\documentclass[onecolumn,authoryear]{els-mrw}

\usepackage[T1]{fontenc}
\usepackage{amsmath,amssymb,amsfonts,amsthm,makeidx,graphicx}
\usepackage{txfonts}
\usepackage{helvet}
\usepackage[symbol]{footmisc}

\begin{document}

\chapter{Ultraluminous X-Ray Binaries}\label{chap1}

\author[1]{Kristen C. Dage\footnote{NASA Einstein Fellow}}
\author[2,3]{Konstantinos Kovlakas\footnote{ICE Postdoctoral Fellow}}%

\address[1]{\orgname{Wayne State University}, \orgdiv{Department of Physics \& Astronomy}, \orgaddress{666 W. Hancock St. Detroit, MI, 48201, USA}}
\address[2]{\orgname{Institute of Space Sciences (ICE), CSIC}, \orgaddress{Campus UAB, Carrer de Can Magrans s/n, E-08193, Barcelona, Spain}}
\address[3]{\orgname{Institut d'Estudis Espacials de Catalunya (IEEC)}, \orgaddress{Edifici RDIT, Campus UPC, 08860 Castelldefels (Barcelona), Spain}}

\articletag{Chapter Article tagline: update of previous edition,, reprint..}

\maketitle

\begin{glossary}[Key points]
\term{Accretion}: gradual accumulation of matter on to a nearby object \\
\term{Active Galactic Nucleus}: a compact region in the center of a galaxy, emitting radiation due to accretion on to a supermassive black hole \\
\term{Compact Object}: a supermassive black hole or a gravitationally collapsed star, either white dwarf, black hole or neutron star \\
\term{Donor Star}: the star from which the material is being accreted on to another star \\
\term{Geometric Beaming}: non-isotropic radiation, due to the geometry of the accretion disk \\
\term{Eddington Limit}: a theoretical limit of the maximum luminosity that an object can emit \\
\term{Magnetar}: a neutron star with an extremely high (B$ \gtrsim 10^{13}$G) magnetic field \\
\term{Ultraluminous X-ray Source}: an X-ray source with apparent X-ray luminosity exceeding $10^{39}$ erg/s \\
\term{X-ray Binary}: a system exhibiting X-ray emission due to accretion from a donor star on to a compact object \\
\term{X-ray Luminosity}:  the luminosity of an object in the X-ray part of the spectrum, typically 0.5-8.0 keV (although different ranges are used depending on the context and observational facility)

\end{glossary}

\begin{glossary}[Nomenclature]
\begin{tabular}{@{}lp{34pc}@{}}
BH & Black hole \\
HLX & Hyperluminous X-ray source\\
HMXB & High mass X-ray binary\\
IMBH & Intermediate mass black hole \\
LMXB & Low mass X-ray binary\\
$L_X$ & X-ray luminosity \\
NS & Neutron star \\
UCXB & Ultra-compact X-ray binary\\
ULX & Ultraluminous X-ray source\\
XRB & X-ray binary
\end{tabular}
\end{glossary}

\begin{abstract}[Abstract]

Ultraluminous X-ray binaries have challenged our assumptions of extreme accretion rates in X-ray binaries, and impact other subfields of astronomy, such as cosmology, gravitational wave sources and supernov\ae{}. Our understanding of ULXs has changed tremendously over the last 35 years, and we now know that ULXs can be powered by accreting neutron stars as well as black holes, and can be found in a wide range of stellar environments. In this chapter, we introduce the observational techniques used to discover and characterize ULXs, and discuss our current understanding of their unique accretion physics and formation channels. 
\end{abstract}

\section{Introduction}\label{chap1:sec1}
Ultraluminous X-ray Sources (ULXs) are traditionally defined as extragalactic non-nuclear X-ray sources whose X-ray luminosity exceeds the Eddington Limit for a ten solar mass black hole, ~$10^{39}$ erg/s. Their non-nuclear position criterion is used to avoid confusion with active galactic nuclei which are typically located at the centers of their host galaxies, and the X-ray luminosity threshold make them brighter than those usually produced by accreting stellar mass compact objects.

The X-ray emission is produced by a compact object (black hole or neutron star) accreting matter from a donor star.  The Eddington Limit was originally derived by Sir Arthur Eddington in 1926. He began the derivation by combining the hydrostatic equation and the equation of radiative equilibrium, assuming that the radiation is both spherically symmetric, and flowing steadily.

This line of thought can also be applied to accreting black holes to determine the luminosity limit, beyond which the radiation pressure would blow away the material drawn in towards the compact object due to  gravity. This is derived in Chapter 16 of High Energy Astrophysics Vol 2 by \cite{longair}. Here, Thomson Scattering (with cross-section $\sigma_T$) is assumed, so the outward force on the electrons is 

\begin{equation}
f=\sigma_T L/4\pi r^2c,    
\end{equation}
where $L$ is the luminosity, $r$ is radius and $c$ is the speed of light. The gravitational force acting on an electron-proton pair is:  
\begin{equation}
    f_{\rm grav}=\frac{GM}{r^2}(m_p+m_e) \approx \frac{GMm_p}{r^2},
\end{equation}
where $m_p$ and $m_e$ are the mass of a proton and an electron respectively, and $G$ is the gravitational constant.
Consequently, by requiring $f=f_{\rm grav}$ for the maximum luminosity,
\begin{equation}
    \frac{\sigma_T L}{4 \pi r^2 c} = \frac{GMm_p}{r^2}.
\end{equation}
The initial form of the Eddington Limit can now be written as an inequality:
\begin{equation}
    L < \frac{4\pi GMm_pc}{\sigma_T} \equiv L_{\rm Edd}.
\end{equation}
This can be further modified for black holes via the Schwarzschild radius, $r_g=2GM/c^2$:
\begin{equation}
    L_{\rm Edd}<\frac{2\pi r_g m_p c^3}{\sigma_t} \approx 1.3 \times 10^{38} \left (\frac{M}{\rm M_\odot} \right) \textrm{erg/s}.
\end{equation}

It is important to stress that a number of assumptions have gone into the derivation of the Eddington Limit, such as the assumption of spherically symmetric accretion, a steady accretion rate, and material which is dominantly hydrogen and fully ionized. \cite{fkr}'s \textit{Accretion Power in Astrophysics} gives supernov\ae{} as one example where the Eddington Limit is dramatically exceeded by many orders of magnitude. However as they note, the Eddington Limit still provides a useful benchmark, as indeed the population of Galactic and Magellanic Cloud X-ray binaries tends to abide by the Eddington Limit \citep{longair}.

The Eddington Limit generally implies that if we observe a source with an X-ray luminosity greater than 10$^{39}$ erg/s, the compact object is more massive than a ten solar mass black hole, and could possibly provide an observational means to identify objects from the elusive class of intermediate mass black holes (100-10$^5$ $\rm M_\odot$ black holes). Early work on ULXs, such as \cite{colbert1999} suggested them as natural IMBH candidates. However, the rate of observed ULXs were already at odds with, and far outpaced, predictions of the number of IMBHs that would form, and it was suggested that the majority of ULXs were in fact systems with accreting stellar mass black holes \citep[see][and refererences therein]{2009MNRAS.397.1836G}.

The number of assumptions that have gone into the derivation of the Eddington Limit provide many opportunities for a stellar mass compact object to produce, or appear to produce X-ray luminosities above its own Eddington Limit. \cite{2001ApJ...552L.109K} have suggested that ULXs may be geometrically beamed (i.e. the flux is only being emitted over some fraction of a sphere, instead of the entire sphere), and may represent a short lived evolutionary stage of the larger population of X-ray binaries, rather than a separate class of objects. The geometrical beaming, if not accounted for, results in higher luminosity estimates which typically assume isotropic emission. A ten solar mass black hole with a mild beaming factor of 0.1 or less will still produce observed luminosities above the nominal Eddington Limit, without being significantly different from a typical X-ray binary system. However, as we will discuss further, a number of observational studies have suggested that geometric beaming, while necessary to explain extreme X-ray luminosities, cannot solely account for all ULX systems and moderately or highly super-Eddington accretion is also necessary. 

While the garden-variety class of ULXs can be interpreted by some combination of geometric beaming or super-Eddington accretion onto a stellar mass compact object, the class of hyper-luminous X-ray sources ($L_X>10^{41}$ erg/s) represent the most likely IMBH candidates. A notable example includes HLX-1 \citep{2009Natur.460...73F}, which is discussed further in Section 1.1.3.

Our understanding of ULXs changed significantly in 2014, when pulsations$-$from the surface of a neutron star$-$were discovered from the ULX M82 X-2 \citep{2014Natur.514..202B}. This complicated our understanding of ULX systems, as there is not yet a clear metric to distinguish between NS ULXs and BH ULXs \citep{2018ApJ...856..128W} for the vast majority of the ULX population. Since the Eddington limit of NSs is ${\approx}2{\times}10^{38}\rm\,erg/s$, the discovery of NS ULXs also made highly super-Eddington accretion necessary for at least a part of the population.

The nature of ULXs is thus extremely mysterious, as there is not yet a systematic way to determine whether the X-rays are produced by a neutron star, a stellar mass black hole, or an intermediate mass black hole. ULXs therefore may provide constraints on the population of intermediate mass black holes, but they also provide constraints on how compact objects can accrete under some of the most extreme environments and accretion scenarios. 

The class of ULXs is therefore expected to contain subpopulations with differences in the type of accretor/donor, evolutionary paths, and underlying physical mechanisms. IMBHs in ULXs are sub-Eddington sources that may connect current black hole populations to those that were used as seeds of supermassive black holes in the early Universe. Stellar-mass black holes in ULXs must be near- or super-Eddington accretors, providing an avenue to study accretion physics at extreme mass-transfer rates, with applications also to active galactic nuclei. Finally, neutron-star ULXs are excellent laboratories for probing highly super-Eddington accretion, and most importantly, the effect of magnetic fields in accretion flows or the possible connection to magnetars. 

\section{Observations of ULXs}\label{chap1:subsec1}

Since the the first experiments from Riccardo Giacconi, and the first X-ray sky survey with the \textit{Uhuru} observatory in the 1970s, our understanding of bright X-ray sources has come a long way. The \textit{Einstein Observatory} in 1978 enabled the study of extragalactic X-ray sources, culminating in the eventual classification of ULXs \citep{1989ARA&A..27...87F}. Early studies of ULXs were enabled by the \textit{Röntgensatellit (ROSAT) }in the 1990s, but it was not until the launch of the \textit{Chandra X-ray Observatory }in 1999, that the discovery and characterization of ULXs happened on a larger scale. The later launches of the \textit{High Throughput X-ray Spectroscopy Mission and the X-ray Multi-Mirror Mission (XMM-Newton) }and the \textit{Nuclear Spectroscopic Telescope Array (NuSTAR)} allowed for a broad band (0.1 to 79 keV) study of ULXs, and enabled their timing properties to be studied. 

We note that a number of Galactic X-ray binaries (e.g. GRS 1915) will often surpass their own Eddington Limits in outburst, however, for the purposes of this chapter, we will mainly focus on sources with sustained high X-ray luminosities, which so far are only found outside the Milky Way.

\subsection{What X-rays tell us}
Beyond localization and X-ray luminosity, we can learn a number of useful information from X-ray observations. Uniquely, a given X-ray observation  contains the following information: the position of the X-ray source on the sky, the number and energy of the photons collected, and the relative time of photon arrival.

 \subsubsection{X-ray spectroscopy}
\begin{figure}
    \centering
    \includegraphics[width=8in]{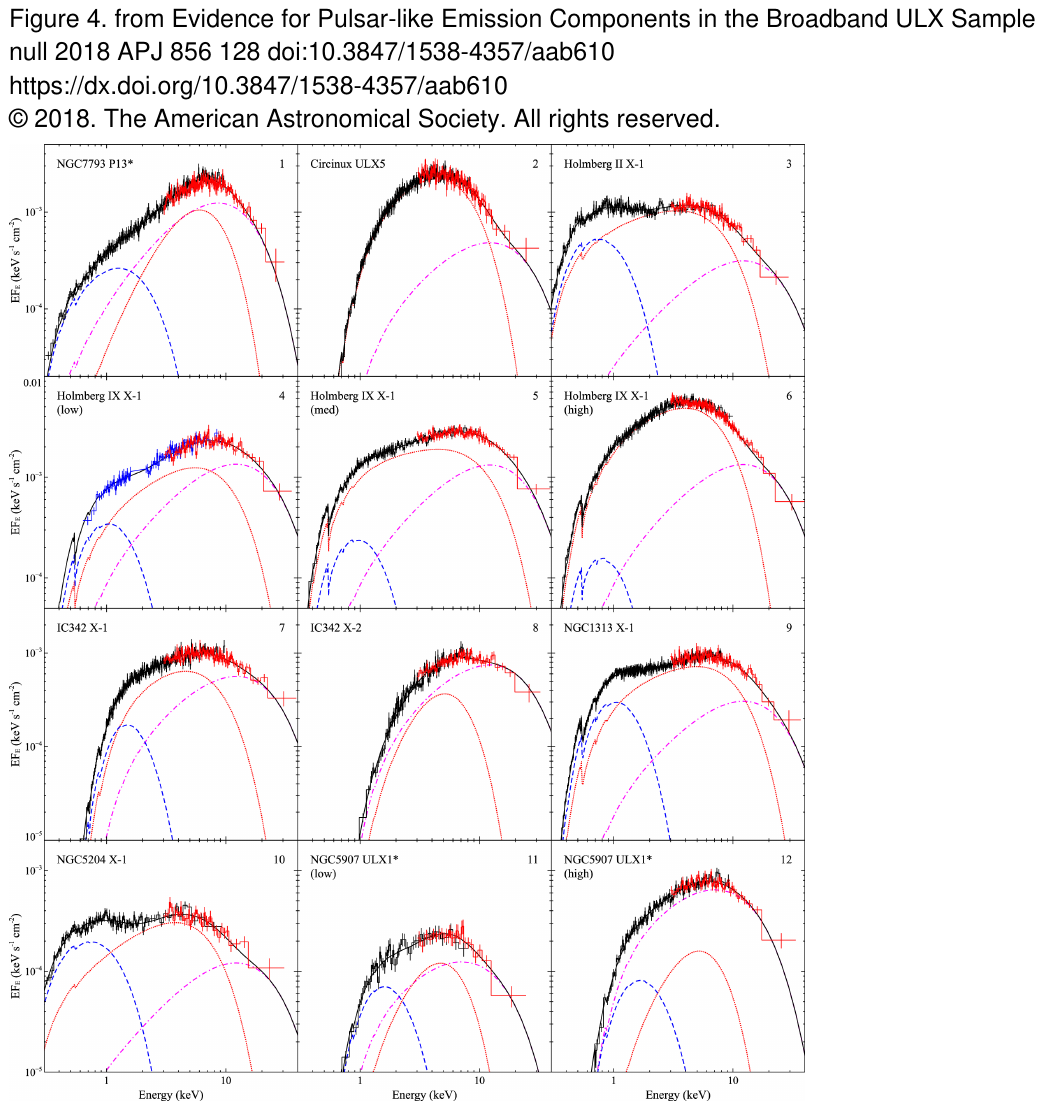}
    \caption{Figure 4 from \cite{2018ApJ...856..128W}. The black and blue solid lines represent lower energy observations from \textit{XMM-Newton} and \textit{Suzaku} respectively. The red solid lines are high energy observations from \textit{NuSTAR}. The blue dashed lines and dotted red lines represent thermal emission models, and the magenta dash-dot lines represent the power-law model for a high energy excess. "Low", "med" and "high" indicate the relative fluxes for the same sources.  This result shows that all of the ULXs in the sample show this high energy excess, regardless of whether or not they are neutron stars.  © AAS. Reproduced with permission.}
    \label{fig:Walton18}
\end{figure}

 The physical properties of the accretion mechanism can be probed via X-ray spectroscopy. The distribution in the number of photons observed at a given energy (i.e., spectral energy distribution), is the result of the convolution of the emitted spectrum with the known properties of the X-ray telescope, and therefore can be used to understand the underlying physical emission mechanisms. For example, thermal emission from an accretion disk can be modeled as a blackbody component with a peak in lower energy X-rays. Non-thermal emission will be modeled as a power-law component to fit any photons observed at higher energies. These models can be combined with components to model absorption (both intrinsic absorption from the source, or to account for X-rays absorbed via known interstellar medium in the Galactic line-of-sight between the X-ray detector and the object's position). Examples of these are shown in Figure \ref{fig:Walton18}; the solid lines represent the data obtained from X-ray telescopes, and the dashed and dash dotted lines represent the various model components used to fit the data. 

 A systematic study of ULX spectra by \cite{2009MNRAS.397.1836G} found that the X-ray spectral shape can set ULXs apart into two classes. A subset of ULXs, typically those with comparatively lower X-ray luminosities (i.e. closer to $10^{39}$ erg/s) have spectral features which are very similar to the brightest stages of Galactic X-ray binaries in outburst. This suggests that there is overlap between the brightest Galactic X-ray binaries and the faintest extragalactic ULXs. A larger number of ULXs, however, show very different spectral features.  These are characterized by a thermal disk component with a power-law tail, which shows a breaking feature near 2\,keV. These features are not observed in any Galactic sources. \cite{2009MNRAS.397.1836G} suggested the breaking feature near 2\,keV could be an observational signature of strong winds covering the inner accretion disk. Thus, this work became the basis for identifying a new ``ultraluminous state'' associated with super-Eddington outflows. 

 After the discovery of pulsating ULXs in 2014, the natural next question was whether the spectral shapes of pulsating ULXs were vastly different than the spectra of the general population. \cite{2018ApJ...856..128W} performed a broadband study of nine ULXs,  using data from simultaneous observations by \textit{XMM-Newton} and \textit{Suzaku}, which covered the low energy X-rays, and \textit{NuSTAR} which observed the high energy X-rays. Two of this sample, NGC 7793 P13 and NGC 5907 ULX1, were known pulsating ULXs, whereas the nature of the rest remains unknown.

 As shown in Figure \ref{fig:Walton18}, the sources were fit with black body disk models, which represent the thermal continuum flux produced by a broadened accretion disk. All sources had an excess of high-energy photons beyond the thermal disk models, which was modeled with a power-law. For the two pulsating ULXs, \cite{2018ApJ...856..128W} performed phase-resolved spectroscopy by extracting the spectral information that was produced when the pulses were on in the system. In both instances, when the pulses were on, the high energy component was present, and thus a possible physical interpretation is that the high energy excess is associated with magnetically collimated accretion columns of the neutron star. The prevalence of this high energy emission component in the spectra of the ULXs sampled by \cite{2018ApJ...856..128W} suggests, at the very least, that broadband X-ray spectroscopy does not, for the moment, provide a satisfactory metric to separate black hole ULXs from neutron star ULXs, or implies that a significant number of ULXs are neutron stars.

High resolution X-ray spectroscopy, including studies performed with the Reflection Grating Spectrometer on \textit{XMM-Newton}, can provide evidence for the presence of ultra-fast outflows in several ULXs \citep[e.g.,][]{2018MNRAS.479.3978K}. 

Several studies of ULXs, using high signal-to-noise \textit{Chandra} spectra, have found significant absorption lines in the spectra. These are suggested to be representative of a cyclotron resonance scattering feature (CSRF) from a neutron star. Should this be the case, the line energies and features could imply very high magnetic fields, possibly as high as $\sim 10^{15}$ Gauss. High magnetic fields will reduce the electron scattering cross section ($\sigma_T$ in Equation 4), and thus reduce the limiting effects that radiation pressure has on the maximum luminosity. This offers an alternate explanation to how neutron stars can surpass the Eddington Limit, beyond solely geometric beaming.

\subsubsection{X-ray timing} One of the key results to come out of X-ray timing of ULXs was the discovery of coherent pulsations from ULXs observed by \textit{NuSTAR}, beginning with \cite{2014Natur.514..202B}. We will refer to these as pulsating ULXs. The first of these, M82 X-2 was discovered to have 1.37 second pulsations. These are caused by matter falling onto the magnetized surface of a neutron star, producing collimated emission spinning at the rate of the neutron star's rotation. A follow-up study of M82 X-2's pulsations revealed pulse drop-outs. Thus, while the observation of pulsations can rule in a neutron star primary, the absence of pulsations does not rule out a neutron star. Since the initial discovery of pulsations in M82 X-2,  pulsations have been observed in a total of seven ULX systems, but thanks to pulse transience, there is not yet a clear method to distinguish black hole accretors from neutron star ULXs via timing techniques. 

Other timing signatures can appear over several time scales, and can reveal key clues about the nature of the ULX binary makeup. The orbital period, i.e. observational periodic signatures caused by the orbit of the compact object and the donor star, has timescales on the order of days, can help to broadly constrain properties of the donor star and binary separation. In some cases, super-orbital periods (periods longer than the system's orbital period) are detected in X-rays. These super-orbital periods have been connected to the geometry of the accretion disk, and may be explained by a warped, precessing accretion disk, where the warp of the disk rotates and obscures the system. 

Short term, but not necessarily periodic, flaring behavior can also be observed in a few very unique systems. Thus, ULXs are not only extreme in their observed X-ray luminosities, but also demonstrate transient behavior, making them intriguing in the time domain \citep[see][and references therein]{2023NewAR..9601672K}.

 \subsection{Multi-wavelength followup to ULXs}
While the canonical definition of a ULX is mainly based on its X-ray luminosity, and X-ray techniques have been used to explore properties of ULXs in a systematic manner, information from other wavelengths can be crucial to illuminate key aspects of ULX systems. Optical and infrared followup can provide constraints on the nature of donor stars in ULX binaries, whereas radio, optical and ultraviolet can inform on the unique accretion processes present in the observed systems. The implications of mutiwavelength follow-up on interpreting the nature of the ULX binary, and the challenges in distinguishing the effect of environment versus radiation produced by extreme accretion are discussed below.

  \subsubsection{Signatures of super-Eddington accretion, and bubbles}

 Due to the relative paucity of sensitive ultraviolet instruments, it is often more difficult to perform ultraviolet follow up of ULXs. An additional issue is that ultraviolet signatures from accretion and from the environment can become entangled, making it difficult to determine whether the ultraviolet emission is due to nearby supergiant stars, or from the reprocessing of the X-ray in the outer accretion disk, resulting in their downscattering to the ultraviolet part of the spectrum.

Many optical spectroscopic studies of ULXs have revealed extended line emission nebul\ae{} in regions up to 500 parsec wide. In many of these systems, the presence of nebul\ae{} have been used to argue against geometric beaming for the origin of the bright X-ray luminosity. Because the optical emission is associated with a nebula, it cannot be beamed, due to the large size scale. If the X-rays are beamed, then there should thus be instances of optical nebul\ae{} with no bright X-ray counterpart (i.e. the X-rays are beamed away from us). For now, the observations are not in favor of geometric beaming as the sole explanation for the high X-ray luminosity of all ULXs. 

As discussed in \cite{2023NewAR..9601672K}, there are no widespread studies of radio counterparts to ULXs, with only a handful of them found associated with radio sources. Excitingly,  radio bubbles have also been discovered in many systems. While their energetics are poorly understood as of yet, they serve to paint an intriguing picture of both the unique accretion in these systems, as well as their origins and evolution. The horizons for discovering and characterizing the radio and optical counterparts in ULXs, particularly bubbles, will be significantly expanded in the future with the advent of upcoming optical and radio survey telescopes, such as the \textit{Vera C. Rubin Observatory}, and the \textit{Square Kilometre Array}, and planned future telescopes such as the \textit{next-generation Very Large Array}.

\subsubsection{ULX donor stars and environments}
While the majority of ULXs tend to be discovered in spiral galaxies with intense star-formation activity, they are found in all kinds of galaxies \citep[e.g.][]{Kovlakas20}. A detailed study of their optical counterparts$-$and thus their environments$-$ can provide key information regarding the nature of the ULX. 

While many optical and NIR studies have searched for counterparts to ULXs, there are multiple issues with identifying the correct donor star in the ULX binary. The first issue is crowding; there are often multiple optical counterparts within the positional uncertainty of the X-ray source, and thus it is not possible to know for certain which star is the actual counterpart. The second issue is that owing to their large distances, only the most luminous stars will be detected. If the actual donor star is too faint, it will not be picked up in the observation (e.g., simulations by \citealt{2017ApJ...846...17W} suggest that the majority of ULXs would have red giant companions, which many observations would not be sensitive to). Of the hundreds of ULXs identified, only a small handful have confident classifications of the donor stars.

A significantly smaller number of ULXs are observed with globular cluster counterparts. However, these sources offer an important difference $-$globular clusters are older populations of stars, and the donor star must be of mass lower than the Sun. Furthermore, the ULX binary is likely to be a product of dynamical formation,  where the compact object and the donor star can be of different age and evolved separately, and later formed due to interactions with other stars in the cluster. In one case, optical spectroscopy has provided strong evidence for a white dwarf donor star for the ULX, implying that the system is an ultra-compact X-ray binary (UCXB) \citep{Dage19}.  Most UCXBs are expected to undergo a phase of super-Eddington accretion early in their evolutionary history, so it is also possible to place some constraints on the age, mass transfer rate and orbital period of the system.

\subsection{Individual ULXs of note}
ULXs can be studied both as a population, and as individuals. In depth, long-term, multiwavelength studies of ULXs can reveal a diversity of behavior within the class. Below we present a few individual ULXs which highlight the numerous different combinations of environments, compact objects and donor stars that can produce a ULX.

\subsubsection{SS433}
SS433, first discovered in 1977, is considered an archetypal ULX. SS433 does not technically meet the nominal definition of a ULX, as its \textit{observed} X-ray luminosity does not exceed the Eddington Limit. However, its relativistic jets, which emit from X-ray to radio, and supercritical mass transfer rate strongly suggest that SS433 is a ULX, but viewed with an edge-on orientation. Because SS433 is relatively nearby in the Milky Way, it provides multiwavelength observations that offer a lens to consider properties of more distant ULXs. SS433 may also offer a window into the hidden population of ULXs that we are unable to detect because of the distance and orientation.

\subsubsection{HLX-1}
The hyper-luminous X-ray source HLX-1 in the galaxy ESO 243-49 is often considered to be one of the best IMBH candidates. It was observed with a maximum X-ray luminosity of $10^{42}$ erg/s at a distance of 95 Mpc, and is strongly variable. The maximum luminosity is three orders of magnitude higher than the Eddington Limit for a ten solar mass black hole. HLX-1 is offset significantly (3.3kpc) from the central nucleus of the host galaxy, and thus cannot be the central supermassive black hole. Observations by \cite{2014MNRAS.437.1208F} suggest that HLX-1 is hosted by a star cluster, although there is a degeneracy as to whether it is hosted by a young stellar population ($\sim 10^6$ years), or an older stellar population ($\sim 10^{10}$ years). However, an alternate interpretation is that HLX-1 was the central massive black hole of a small galaxy which was tidally stripped by ESO 243-49. If so, then it suggests that HLX-1 is not an IMBH with a unique formation channel, but rather a petite supermassive black hole \citep{2020ARA&A..58..257G}

\subsubsection{RZ2109}
RZ2109 is a ULX hosted by a confirmed globular cluster in Virgo \citep{2007Natur.445..183M}. Uniquely, the ULX formed, possibly through dynamical interactions, in an old stellar population. Optical spectroscopy of the system reveals [OIII] $\lambda\lambda$  4959,5007 \AA \hspace{0.1cm} emission, which has a broadened profile that varies over many years of study, and strongly implies that the donor star in the system is a carbon-oxygen white dwarf \citep{Dage19}. Because of these properties and the host environment, RZ2109 is very different from the vast majority of ULXs and highlights the diversity of ULX systems.

\begin{figure}
    \centering
    \includegraphics[width=5in]{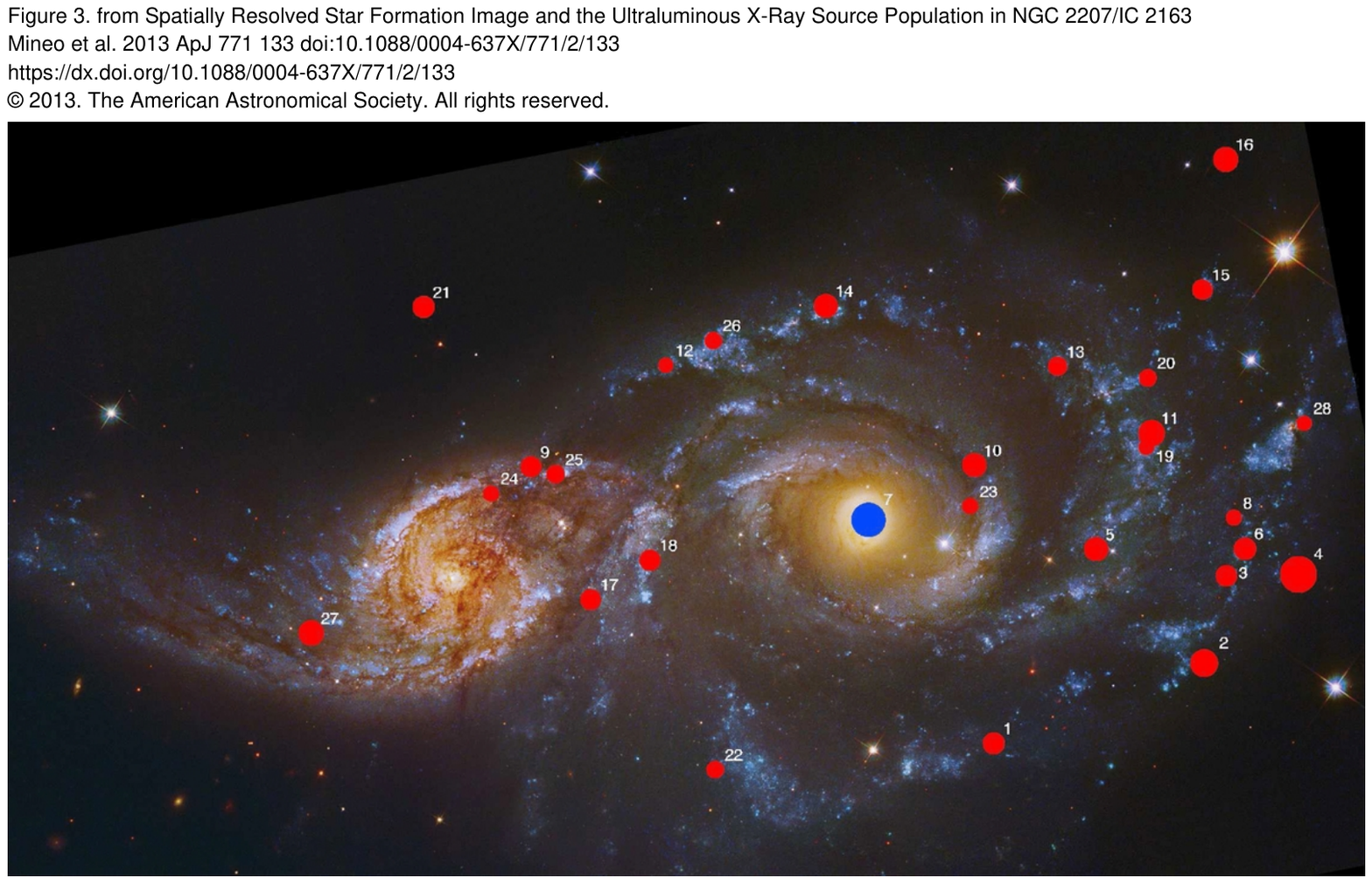}
    \caption{HST image of NGC 2207/IC 2163 from Figure 3 from \cite{2013ApJ...771..133M}. The blue region masks out the X-ray emission from the central region of the galaxy NGC 2207, and the red regions represent the distribution of X-ray binaries detected in the galaxy, with the region size proportional to X-ray luminosity. 21 of the 28 X-ray point sources are ULXs, one of the largest numbers of ULXs per galaxy mass discovered. The population of ULXs can range to large distances from the host galaxy, and for this galaxy are correlated with local star formation density. © AAS. Reproduced with permission.}
    \label{fig:mineo}
\end{figure}

\subsection{Where to find ULXs}

While observations of individual ULXs are critical for understanding their nature and emission mechanisms, only the ones closest to us are relatively well studied. Owning to their rarity (typically 0-2 ULXs per galaxy), the majority of ULXs are hosted in distant galaxies. Consequently, demographic studies and statistical approaches aim at complementing our view of ULXs. Despite the inability to probe the nature of the compact object and the donor stars, the numbers and luminosity distribution of ULXs in different galactic environments allow us to indirectly make conclusions regarding their formation and evolution (see Figure \ref{fig:mineo} for an example of ULX distribution in a galaxy).

ULXs are found in all types of galaxies: spiral, elliptical and irregular. Despite the generally-used and historical definition of ULXs on the basis of their observational appearance as extragalactic and non-nuclear, one would notice that these criteria are used mainly to decrease the contamination in demographic studies from other types of sources. However, super-Eddington stellar mass accretion objects cannot be excluded in the Milky Way. In addition, a non-negligible fraction of ULXs is expected to be hosted at the central regions of galaxies due to their high density in stars and star-forming gas. A nuclear source should in fact be classified as an ULX if the existence of an active galactic nucleus can be excluded, while we should always have in mind the caveat that off-nuclear active galactic nuclei do exist and can possibly contaminate ULX samples.

Demographic studies have shown that ULXs scale with the star-formation rate in spiral and irregular galaxies, as expected from a young stellar population. In addition, their numbers are anti-correlated to the metallicity of the host galaxies. Both these two facts suggest a connection to HMXBs that share the same population trends.
Conversely, in passive galaxies (e.g., ellipticals), the number of ULXs depends on the stellar mass as expected from an old stellar population, hinting at a connection with low-mass XRBs.

With the advent of high-resolution X-ray observatories, it was possible to study ULX populations in hundreds of galaxies in the local Universe. In parallel, all-sky photometric and spectroscopic surveys provided the data to characterize the host galaxies' stellar population parameters. As a consequence, recently it was possible to study the ULX content and their demographic properties in an unprecedented detail. For example, \citet{Kovlakas20} used a sample of more than 600 galaxies and equal in number ULXs, to constrain the rate of ULXs in star-forming galaxies as a function of the star-formation rate, the metallicity, as well as in groups of different morphological classes. The rate of ULXs in passive galaxies was found to not depend linearly on the stellar mass of the host, hinting at star-formation history effects (which are averaged out in star-forming galaxy samples) giving higher specific ULX frequency in low-mass passive galaxies.

All the aforementioned trends are the result of a convolution of the star-formation history of a given sample of galaxies with the formation efficiency of binaries exhibiting powerful accretion phases during their evolution. Consequently, the increasing accuracy of the demographic studies is a promising avenue for constraining accretion theory and binary star evolution via combining binary population synthesis models and cosmological simulations.

\section{The theoretical study of ULXs}\label{chap1:sec4}

The emergence of ULXs from binary stars is a combination of binary star evolution and accretion. Uncertainties in key stellar evolution phases \citep[e.g.,][]{Tauris23}, and the exact physical conditions of accretion disks \citep[e.g.,][]{2001ApJ...552L.109K}, do not allow us to forward model ULX populations in a way that is fully consistent with all their observed properties. However, theoretical models are constantly being updated, and constrained, by ULXs.

\subsection{Modelling super-critical accretion disks}

Understanding the geometry and properties of accretion disks at extreme mass transfer rates is critical to the modeling of ULXs as accreting stellar-mass objects. 

One of the most-cited articles in astrophysics, the seminal work in accretion disk theory of \citet{Shakura73}, also approached super-Eddington accretion, showing that super-critical disks would exhibit significant outflow of matter, puffing up of the inner regions of the disk (`spherization'), resulting in a radiative output of the form:
\begin{equation}
    L_{\rm out} \approx L_{\rm Edd} \left(1 + \ln\dot{m}\right),
\end{equation}
where $L_{\rm Edd}$ is the Eddington-limit, and $\dot{m}$ is the Eddington ratio $\dot{M}/\dot{M}_{\rm Edd}$, i.e. the ratio of the accretion rate ($\dot{M}$) over the rate at which the Eddginton limit is reached ($\dot{M}_{\rm Edd}$). Subsequent efforts towards a more realistic description of accretion disks \citep[e.g.,][]{Poutanen07} added in our knowledge of the physical appearance of ULXs as a function of the accretion rate, and the inclination angle, highlighting the effect of geometrical beaming. The beaming of the radiation is a consequence of the geometry of the accretion disk and is expected to initiate at high accretion rates ($\dot{m}\gg1$). While beaming is not observed directly, its assumed dependence on the accretion rate which also affects the spectrum of the disk (and hence the inferred effective temperature), allowed \citet{King09} to derive the relation
\begin{equation}
    b = \begin{cases}
            1,                    &\dot{m} \leq 8.5 \\
            \left(\dfrac{8.5}{\dot{m}}\right)^2, &\dot{m} > 8.5
        \end{cases},
\end{equation}
using observations of the X-ray luminosity and temperature relation of ULXs. Finally, the isotropic-equivalent luminosity of ULXs can be considered as
\begin{equation}
    L_{\rm iso} = \frac{1}{b} L_{\rm out},
\end{equation}
which can explain the extreme luminosities of observed ULXs, without invoking unusually massive BH accretors, or unknown emission mechanisms.

The analytical and empirical models, are generally describing black hole accretors, but can be generalized to include neutron stars if the effects from the magnetic field are negligible.

The increase in computational power, met with advancements in computational astrophysics in the 2010s allowed the use of three-dimensional general-relativistic radiation magnetohydrodynamic simulations \citep[e.g.,][]{Sadowski16} for the study of ULXs. Such simulations confirm the expectations from analytical studies regarding outflows from the accretion disk, the existence of a thick photosphere, and the beaming of the radiation as it escapes from an optically-thin funnel. 

Owning to their computational complexity, the simulations often exclude physics, have only been used to study a few examples of system configurations, and typically last for very short physical times (of the order of seconds!) As a consequence, the results depend highly on the computational framework and physical assumptions, they cannot be used to extrapolate their results in population synthesis studies, and cannot probe many of the observational aspects of ULXs (e.g., variability).

The discovery of pulsar ULXs with NS accretors motivated radiation magnetohydrodynamical simulations of accretion disks around magnetized neutron stars. Similar characteristics to BH ULXs, such as powerful outflows and geometrical beaming, are found for NS ULXs of low \citep[e.g., $B{<}10^{11}\rm\,G$;][]{Abarca21} and intermediate magnetic fields \citep[e.g., $B{\sim}10^{12}\rm\,G$;][]{Inoue23} when specific conditions are fulfilled (e.g., magnetospheric radius, mass transfer rate). 
Accretion in highly-magnetized ($B{>}10^{13}\rm\,G$) NSs (i.e., magnetars) has not been studied but has been proposed as an alternative explanation of the high X-ray luminosities of pulsating ULXs \citep[e.g.,][]{Eksi15}. However the existence of magnetar ULXs has been contested, mainly on the basis of the fact that magnetars are isolated NSs \citep[e.g.,][]{2023NewAR..9601672K}.

\subsection{Evolution of ULXs and population synthesis studies}

The majority of ULXs are found in actively star-forming galaxies such as spiral and irregular galaxies, with a spatial distribution tracing star formation \citep[see][and references therein]{Kovlakas20}. These are clues to a ULX formation channel from coeval binary systems with high-mass donors in young stellar populations.

The clues from the connection of ULXs with their parent stellar populations, and the X-ray emission mechanisms from accretion are not enough to fully understand how ULXs are formed. 
Binary stellar evolution theory is central in understanding how extreme mass transfer rates can occur in systems with a compact object and a non-degenerate star. 

Binary evolution studies focusing on the X-ray phases of the massive binaries \citep[e.g.,][]{Podsiadlowski03,Rappaport05} illustrated the importance of binary interactions in the evolution of ULXs, and showed that the measured rate of ULXs in local Universe galaxies can be explained by populations of HMXBs with BH accretors.
In addition, binary stellar evolution studies can discover and measure the contribution of different ULX formation channels. As one would expect from the current understanding of ULXs being the natural extension of XRBs at high luminosities, ULX evolution is often studied along with the rest of the XRBs, or employing the same computational frameworks \citep[e.g.,][]{Misra20}. Such studies have significantly improved our understanding of how extreme mass transfer rate can emerge in short-lived phases in the lives of the stars via Roche-lobe overflow, wind loss, or decretion disks \citep[in Be-XRBs; see][and references therein]{2023NewAR..9601672K}.

After the discovery of pulsar ULXs \citep{2014Natur.514..202B}, special focus was given to the understanding of the evolutionary scenarios for the formation of ULXs and their subpopulations, as well as the ratio of NS/BH accretors in the systems. The large-scale ULX population synthesis study of \citet{2017ApJ...846...17W} identified multiple system configurations, on the basis of the compact object (NS or BH) and companion type (main-sequence, red giant, Hertzsprung gap stars), and characterized the age of their parent stellar populations. In addition, it highlighted the importance of the metallicity in the fraction of NS ULXs over the total ULX population as a function of the age. The known existence of ULXs in elliptical galaxies indicates the possibility of ULXs arising from old stellar populations, and therefore with low-mass companions as found in population synthesis studies. While it is impossible to measure the age of the parent populations of ULXs, indirect evidence comes via demographic studies: \citet{Kovlakas20} found that the dependence of the number of ULXs in passive galaxies on the host galaxy stellar mass follows the expectations from the \citet{2017ApJ...846...17W} models and average star-formation histories of elliptical galaxies.

The above studies focus on the secular evolution of binary systems, but it possible that interacting binaries can form via gravitational capture in dense stellar populations, such as globular clusters and central regions of galaxies. The existence of ULXs in globular clusters \citep[e.g.,][]{Dage19} challenges a unified formation channel of ULXs. Since elliptical galaxies tend to contain higher numbers of globular clusters, it is possible that a significant fraction of ULXs in passive galaxies can be attributed to this subpopulation. The known dependence of galaxy-wide X-ray luminosity to the globular cluster occupation fraction \citep[e.g.,][]{Lehmer20} provides another hint, but to the date, there is no theoretical study of ULX formation in globular clusters.

\subsection{Connection to other sources and cosmology}

The properties of ULXs make them strong candidates for a variety of astrophysical phenomena. Despite their rarity, they originate from stellar populations and can be found in a wide range of environments. It would be safe to assume that any galaxy with mass above $10^{7}\,\rm M_\odot$ hosted at least one ULX at some point of its life. These ULXs, with their powerful emission and outflows, have affected their surrounding via shocks or ionizing radiation. Despite being less luminous than other energetic sources (e.g., supernov\ae{}, gamma-ray bursts, etc.), ULXs are persistent source and therefore have prolonged effects that makes them observable at high numbers in the local Universe. Having also a wide diversity, such as NS or BH accretor, type of companion, accretion mechanism, they are linked to a multitude of types of astrophysical sources, either at the same evolutionary stage, or as progenitors and successors of other stellar sources.

ULXs being predominantly young sources and particularly abundant in low-metallicity environments, are expected to be teeming in the early Universe right after the first generation of stars. These makes them strong candidates to the heating and re-ionization of the Universe from extreme ultraviolet and shoft X-ray radiation \citep[see][and references therein]{Garofali24}, as well as in explaining emission lines in local Universe galaxies \citep[e.g.,][]{Simmonds21}. Constraints on the ULX content in the early Universe are not only informing us about the X-ray radiation background, but can point at new directions in understanding early star-formation and stellar evolution at near-zero metallicity environments.

ULXs have also been proposed as sub-classes of other astrophysical sources. For example, \citet{Sridhar21} proposed a mechanism though which ULXs can produce periodic fast radio bursts, complementary to alternative scenarios that are invoked to explain the general population of fast radio bursts.

Population synthesis studies often connect different types of sources on the basis of their evolutionary stage. For example, \citet{Mondal20} show that a significant fraction of double compact object mergers is expected to have passed through an ULX phase. Consequently, we may access important information about the nature and evolution of gravitational-wave sources by studying their progenitors through ULXs as their local-Universe analogs.

Finally, the combination of demographics of ULXs and population synthesis models, can put constraints on binary stellar evolution physics, and accretion theory at extreme accretion rates, which are relevant to a host of areas of research in astrophysics and cosmology, such as active galactic nuclei, supernov\ae{}, and gravitational-wave sources.

\section{Conclusions}\label{chap1:sec5}

ULXs are challenging our understanding of X-ray binaries, and providing an avenue towards greater understanding of the role that ULX binaries may play in other fields of astronomy. ULXs span a wide variety of environments, from star forming regions in spiral galaxies to dense stellar clusters in the outskirts of elliptical galaxies. They are a zoo of diverse objects, ranging from confirmed neutron star primaries to intermediate mass black hole candidates, and their donor stars span from red giant stars to white dwarf stars.  ULXs are challenging to model theoretically, but demographic studies and binary formation simulations do not suggest a unified formation mechanism. The wide diversity of ULXs links them to a number of other fields in astronomy, from cosmology to gravitational waves. 

\begin{ack}[Acknowledgments]
.The authors thank Dominick Walton, and Stefano Mineo for permission to use figures from their work in this chapter. KCD thanks Wasundara Athukoralalage, Ed Brown, Jim Pringle and McKinley Brumback for their help, as well as her local library and MeLCat, the Michigan eLibrary Catalog. KCD acknowledges support for this work
provided by NASA through the NASA Hubble Fellowship grant
HST-HF2-51528 awarded by the Space Telescope Science Institute, which is operated by the Association of Universities for Research in Astronomy, Inc., for NASA, under contract NAS5–26555. KK is supported by a fellowship program at the Institute of Space Sciences (ICE-CSIC) funded by the program Unidad de Excelencia Mar\'ia de Maeztu CEX2020-001058-M.
\end{ack}


\newcommand*\aap{A\&A}
\let\astap=\aap
\newcommand*\aapr{A\&A~Rev.}
\newcommand*\aaps{A\&AS}
\newcommand*\actaa{Acta Astron.}
\newcommand*\aj{AJ}
\newcommand*\ao{Appl.~Opt.}
\let\applopt\ao
\newcommand*\apj{ApJ}
\newcommand*\apjl{ApJ}
\let\apjlett\apjl
\newcommand*\apjs{ApJS}
\let\apjsupp\apjs
\newcommand*\aplett{Astrophys.~Lett.}
\newcommand*\apspr{Astrophys.~Space~Phys.~Res.}
\newcommand*\apss{Ap\&SS}
\newcommand*\araa{ARA\&A}
\newcommand*\azh{AZh}
\newcommand*\baas{BAAS}
\newcommand*\bac{Bull. astr. Inst. Czechosl.}
\newcommand*\bain{Bull.~Astron.~Inst.~Netherlands}
\newcommand*\caa{Chinese Astron. Astrophys.}
\newcommand*\cjaa{Chinese J. Astron. Astrophys.}
\newcommand*\fcp{Fund.~Cosmic~Phys.}
\newcommand*\gca{Geochim.~Cosmochim.~Acta}
\newcommand*\grl{Geophys.~Res.~Lett.}
\newcommand*\iaucirc{IAU~Circ.}
\newcommand*\icarus{Icarus}
\newcommand*\jcap{J. Cosmology Astropart. Phys.}
\newcommand*\jcp{J.~Chem.~Phys.}
\newcommand*\jgr{J.~Geophys.~Res.}
\newcommand*\jqsrt{J.~Quant.~Spectr.~Rad.~Transf.}
\newcommand*\jrasc{JRASC}
\newcommand*\memras{MmRAS}
\newcommand*\memsai{Mem.~Soc.~Astron.~Italiana}
\newcommand*\mnras{MNRAS}
\newcommand*\na{New A}
\newcommand*\nar{New A Rev.}
\newcommand*\nat{Nature}
\newcommand*\nphysa{Nucl.~Phys.~A}
\newcommand*\pasa{PASA}
\newcommand*\pasj{PASJ}
\newcommand*\pasp{PASP}
\newcommand*\physrep{Phys.~Rep.}
\newcommand*\physscr{Phys.~Scr}
\newcommand*\planss{Planet.~Space~Sci.}
\newcommand*\pra{Phys.~Rev.~A}
\newcommand*\prb{Phys.~Rev.~B}
\newcommand*\prc{Phys.~Rev.~C}
\newcommand*\prd{Phys.~Rev.~D}
\newcommand*\pre{Phys.~Rev.~E}
\newcommand*\prl{Phys.~Rev.~Lett.}
\newcommand*\procspie{Proc.~SPIE}
\newcommand*\qjras{QJRAS}
\newcommand*\rmxaa{Rev. Mexicana Astron. Astrofis.}
\newcommand*\skytel{S\&T}
\newcommand*\solphys{Sol.~Phys.}
\newcommand*\sovast{Soviet~Ast.}
\newcommand*\ssr{Space~Sci.~Rev.}
\newcommand*\zap{ZAp}
\bibliographystyle{Harvard}
\bibliography{reference}

\end{document}